 \documentclass[preprint,review,12pt]{elsarticle}

 \usepackage{graphicx}

\usepackage{amssymb}
\usepackage{epstopdf}
\usepackage{latexsym}

\journal{Physics Letters B}

\begin{document}

\begin{frontmatter}



\title{Gluodissociation and Screening of $\Upsilon$ States in PbPb Collisions at $\sqrt {s_{NN}}$ = 2.76 TeV}


\author{Felix Brezinski and Georg Wolschin}

\address{ Institut f{\"ur} Theoretische 
Physik
der Universit{\"a}t Heidelberg, 
        Philosophenweg 16,  
        D-69120 Heidelberg, Germany, EU}

\begin{abstract}
We suggest that gluon-induced dissociation and screening of the $\Upsilon(nS)$ states explain
the suppression of the $\Upsilon(2S+3S)$ states relative to the $\Upsilon(1S)$ ground state that has been observed by CMS in PbPb collisions at $\sqrt{s_{NN}}$ = 2.76 TeV at the CERN LHC. The minimum-bias gluodissociation cross sections of the $1S-3S$ states are calculated using a screened Cornell potential and a thermal gluon distribution.
The $3S$ state dissolves due to screening before sizeable gluodissociation occurs, but for the $2S$ and $1S$ states there is an interplay between screening, gluodissociation, and feed-down from the $\chi_b(2P)$
and $\chi_b(1P)$ states. 
Based on a schematic approach, we find that the calculated suppression of the
$\Upsilon(2S)$ and $\Upsilon(3S)$ states relative to $\Upsilon(1S)$ is consistent with the CMS result, but allows for additional suppression mechanisms. The $\Upsilon(1S)$ suppression
through gluodissociation is, however, in good agreement with the CMS data.\\
\end{abstract}


\end{frontmatter}



\newpage

The suppression of quarkonium states is one of the most promising probes for the properties of the quark-gluon plasma (QGP) that is generated in heavy-ion collisions at high relativistic energies. In the QGP the confining potential of heavy quarkonium states is screened due to the interaction of the heavy quark and the antiquark with medium partons and hence,  charmonium and bottomium states successively melt \cite{ms86} at sufficiently high temperatures $T_{{diss}}$ beyond the critical value $T_c\simeq 170$ MeV. 


Charmonium suppression has been studied since 1986 in great detail both theoretically, and experimentally at energies reached at the CERN Super Proton Synchrotron SPS,  BNL Relativistic Heavy-Ion Collider (RHIC) \cite{klusa09,kha07,eta09,ps01}, and CERN LHC \cite{ma11,csi11}. The precise origin is still under investigation, in particular at LHC energies where regeneration due to statistical recombination of $c$ and $\bar{c}$ in the quark-gluon plasma could be relevant, counteracting the $J/\Psi$ dissociation in central collisions and contributing to the measured nearly flat suppression factor as function of centrality for $p_T>0$ \cite{ma11}.

Bottomium suppression is expected to be a cleaner probe. The $\Upsilon(1S)$ ground state with invariant  mass 9.46 GeV 
is strongly bound, the threshold to $B\bar{B}$ decay is at 1.098 GeV. Its lifetime of 1.22$\cdot 10^{-20} $s is about 1.7 times as large as the one of $J/\Psi(1S)$ in elementary collisions. It melts as the last quarkonium in the QGP (depending on the potential) only at 4.10 $T_c$ \cite{won05}, whereas the $2S$ (10.02 GeV) and $3S$ (10.36 GeV) states melt at about 1.6 and 1.2 $T_c$, respectively.
Even at LHC energies the number of bottom quarks in the QGP remains small such that statistical regeneration of the $\Upsilon$ states is unimportant.

 $\Upsilon$ suppression in heavy-ion collisions has recently been observed for the first time both by the STAR experiment at RHIC \cite{hma11}, and by the Compact Muon Solenoid (CMS) experiment at LHC \cite{cha11}. The latter includes an observation of the enhanced suppression of the $2S+3S$ relative to the $1S$ ground state, whereas the $1S$ suppression itself is considered in \cite{chat11} by CMS. 

This result is most likely not due to differences in the direct bottomium production mechanism in $pp$ vs. PbPb collisions since nuclear modification of the parton distribution functions (shadowing) should affect all three states in a similar fashion \cite{chat11}.

In this Letter we investigate the suppression of $\Upsilon(1S), (2S), (3S)$ states at LHC energies due to  screening and gluon-induced dissociation, including feed-down from the $\chi_{b}(1P)$ and $\chi_{b}(2P)$ states. Whereas gluodissociation below $T_c$ is not possible due to confinement, it does occur above $T_c$ where the color-octet state of a free quark and antiquark can propagate in the medium.
The process is relevant below the dissociation temperature $T_{{diss}}$ that is due to Debye screening, and its significance increases substantially with the rising gluon density at LHC energies.  

In the midrapidity range $|y|<2.4$ where the CMS measurement \cite{cha11} has been performed, the temperature and hence, the thermal gluon density is high, and causes a rapid dissociation in particular of the $2S$ and $3S$ states, but also of the $1S$ ground state. At larger rapidities up to the beam value of $y_{beam}=7.99$ and correspondingly small scattering angles where the valence-quark density is high \cite{mtw09}, nonthermal processes would be more important than in the midrapidity region that we are investigating here. 
Thermal gluons will also dissociate the  
 $\chi_{b}(1P)$ and $\chi_{b}(2P)$ states which partially feed the $\Upsilon(1S)$ ground state in elementary collisions \cite{aff00}.


Due to the small velocity $v\ll c$ of the quarks in the bound state, the proper equation of motion for single-particle quarkonium states is the Schr\"odinger equation, with the color-singlet $Q\bar{Q}$ quarkonium potential $V_{Q\bar{Q}}$. 
Reasonable parametrizations of the potential exist that have been tested in detailed calculations of the excited states.

In particular, the Cornell potential \cite{ei75} has string and Coulomb part 
$V_{Q\bar{Q}} = \sigma r - \alpha_{{eff}}/r$,	
where $\sigma \simeq 0.192$ GeV$^2$ \cite{ja86} is the string tension, and $\alpha_{{eff}} =0.471$  
an effective Coulomb-like coupling constant that accounts for the short-range gluon exchange, respectively. 

Although the string contribution to the potential vanishes for light quarkonia in the QGP above $T_c$, it has to be considered at $T>T_c$ for heavy quarkonia that remain initially  confined and are therefore not in thermal equilibrium with the plasma. Hence we maintain the string contribution in an approximate solution of the gluodissociation problem.

The string tension of quarkonium decreases with increasing temperature $T$ in the quark-gluon medium. 
The screened potential can be written as \cite{ja86,kms88,ber08}
\begin{equation}
V(r,T) = \sigma r_D \left[ 1 - e^{- r/r_D}\right]  -
\left[ \frac{\alpha_{{eff}}}{r_D} +\frac{\alpha_{{eff}}}{r}e^{- r/r_D}\right]	\\
\end{equation}
with $r_D(T)$ the Debye radius, $ r_D^{-1} =T\,[4\pi\alpha_s (2 N_c + N_f)/6]^{1/2}$. 
The number of colors is $N_c=3$, the number of flavors in the QGP taken as $N_f=3$, and the strong-coupling constant at the $\Upsilon(1S)$ mass $\alpha_s \simeq 0.2$.
Because of the inverse proportionality of the minimum screening radius that permits a bound state to the heavy-quark mass \cite{ms86}, it is much more difficult to dissolve the $\Upsilon(1S)$ in the quark-gluon plasma through screening  than the $J/\psi(1S)$. 

\begin{figure}
\begin{center}
\includegraphics[width=10cm]{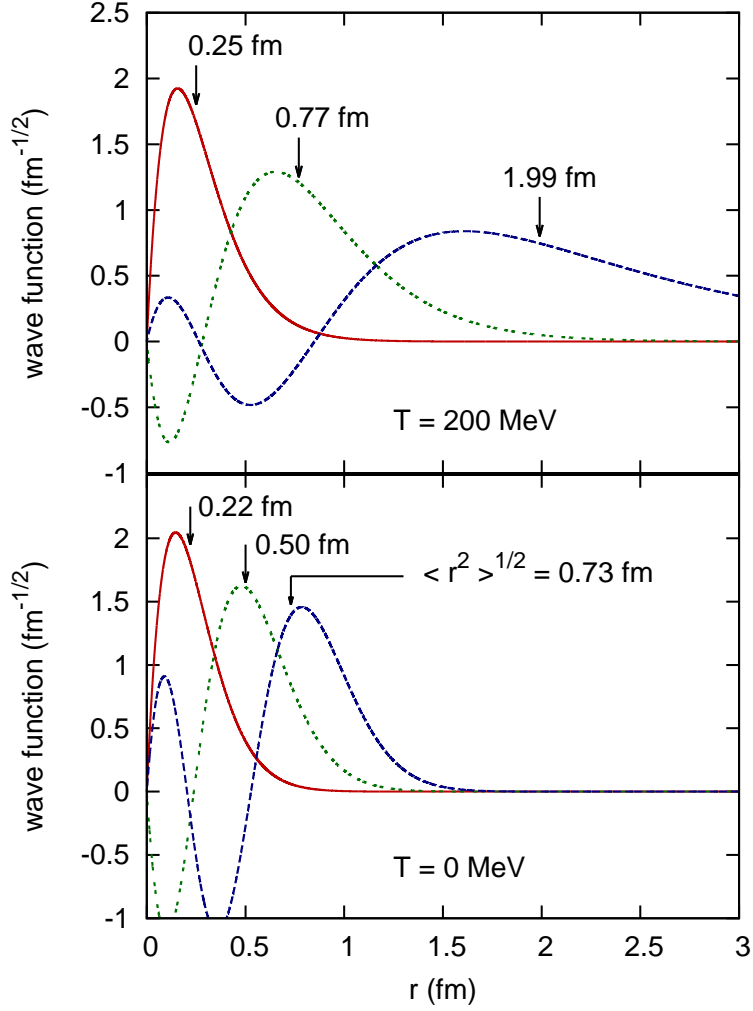}
\caption{\label{fig1} (color online) Radial wave functions of the $\Upsilon (1S), (2S), (3S)$ states (solid, dotted, dashed curves, respectively) calculated in the screened Cornell potential for temperatures $T=0$ MeV (bottom) and 200 MeV (top) with effective coupling constant $\alpha_{{eff}}=0.471$, and string tension $\sigma=0.192$ GeV$^2$. The rms radii $<r^2>^{1/2}$ of the
$2S$ and, in particular, $3S$ state are strongly dependent on temperature $T$, whereas the ground state remains nearly unchanged.} 
\end{center}
\end{figure}

We have calculated the wave functions of the $1S-3S$ states, as shown in Fig.~\ref{fig1} for $T=0$ and 200 MeV. They are almost independent on temperature for the ground state. For the $2S$ state, there is an increase of the rms radius from $0.50$ to $0.77$ fm, whereas for the $3S$ state the rms radius increases from $0.73$ to $1.99$ fm. 

\begin{table}
\begin{center}
\caption{\label{table1}%
Thermally averaged cross sections $<\sigma_{diss}(nS)>$ in mb for the gluodissociation of the $\Upsilon (1S), (2S), (3S)$ states at four different temperatures $T$
and $m_g=0$ in 2.76 TeV PbPb.
The values include screening as described in the text; $2S$ and $3S$ states are screened completely at high $T$.} 
\vspace{.2cm}
\begin{tabular}{lllcr}
\hline\
$T$&$<\sigma_{diss} (1S)>$ &
$<\sigma_{diss} (2S)>$ &
$<\sigma_{diss} (3S)>$\\
(MeV)& (mb)&(mb)&(mb)\\
\hline
     
400&0.094&$-$&$-$\\
300&0.141&0.041&$-$\\
200&0.124&0.465&0.152\\
170&0.080&0.783&0.604\\

\hline
\end{tabular}
\end{center}
\end{table}

Due to the high temperature and ensuing large thermal gluon density reached at LHC energies in the midrapidity region, the most important process next to screening that leads to a suppression of Upsilons at LHC is gluodissociation.
Hence we calculate the gluodissociation cross sections for the $1S-3S$, $\chi_{b}(1P)$, and $\chi_{b}(2P)$ states as functions of the initial impact-parameter dependent temperature in the quark-gluon plasma.
Our calculation is complementary to the solution of a Schr\"odinger equation with an imaginary-valued contribution to the potential \cite{lai07,ber08,bram08,bra11} due to Landau damping of the exchanged gluon as performed in \cite{str11} for the $\Upsilon(1S)$ and the $\chi_b(1P)$ states. 

The leading-order dissociation cross section of the $Q\bar{Q}$ states through dipole interactions with hard gluons ($E1$ absorption of a single gluon) had been derived by Bhanot and Peskin (BP) \cite{bp79}. In an operator product expansion, they calculate the
gluodissociation cross section $\sigma_{{diss}}$ with pure Coulomb-like momentum eigenstates.
This expansion is valid for sufficiently small bound-state radii.
 For an initial gluon of energy $E$ (momentum $p$) the cross section is obtained from the Born amplitude $A_B$ using the optical theorem $Im [A_B(t=0)] = E\sigma_{{diss}}$. 

Modifying the BP approach to approximately account for the confining string contribution, we use the singlet wave functions computed with Eq.(1). Inserting a complete set of eigenstates 
of the adjoint (octet) Hamiltonian 
$-\Delta/m_b+\alpha_{eff}/(8r)$ with eigenvalues $k^2/m_b$ ($m_b\simeq 4.75$ GeV \cite{ja86} the bottom quark mass)
to calculate the dissociation cross sections of the $\Upsilon(1S,2S,3S)$ and the $\chi_b(1P,2P)$ states \cite{br11}, we obtain 
\begin{eqnarray} 
\sigma_{{diss}}^{nS}(E) = \frac{2 \pi^2 \alpha_s E}{9} \int\limits_0^\infty dk \,  \delta\left( \frac{k^2}{m_b} +\epsilon_n-E \right) 
|w^{nS}(k)|^2\quad  
	\label{improvedsigma2}
\end{eqnarray}
with the wave function overlap integral
\begin{equation}
w^{nS}(k)=\int_0^\infty dr \, r \, g_{n0}^s(r) g_{k1}^a(r) 
\end{equation}
for the singlet radial wave functions $g_{n0}^s(r)$ of the $b$ quark, and the adjoint octet wave functions $g_{k1}^a(r)$. 
The binding energy of the $nS$ state is $\epsilon_n$, and the $\delta$ function accounts for energy conservation, $k^2/m_b=E-\epsilon_n$.

For vanishing string tension $\sigma \rightarrow 0$ and the corresponding values of the binding energy $\epsilon_n$,  a pure Coulomb $1S$ wave function, and a simplification in the octet wave function, this expression reduces to the result in \cite{bp79}. We can, however, evaluate it with the full octet wave function to obtain the $\Upsilon (1S)$ dissociation cross section $\sigma_{1S}$ in terms of the BP expression $\sigma_{{BP}}$
\begin{equation}
 \frac{\sigma_{{diss}}^{1S}}{\sigma_{{BP}}} = \frac{\pi z (1+z^2/4)}{\exp{(\pi z)}-1} \left( 1 + \frac{q z}{4} \right)^2 
 \exp{(2 z
\arctan q)}	\label{improvedsigma3}
\end{equation}
with $z=1/(4q)$, and $q=\sqrt{E/\epsilon_1 - 1}$. The rhs approaches 1 for $z \rightarrow 0$, recovering the BP formula. It approaches 0 for $z \rightarrow \infty$, and agrees with the result obtained 
independently by Brambilla et al. in an effective field theory approach in the corresponding limit \cite{bg11,jg11}. Their work also considers the thermal width of heavy quarkonia due to Landau damping, in addition to the break up of a color-singlet bound state into a quark-antiquark pair that is investigated here. In contrast to other assumptions, these authors find that breakup is the leading term as compared to Landau damping \cite{jg11}.

\begin{figure}
\begin{center}
\includegraphics[width=12cm]{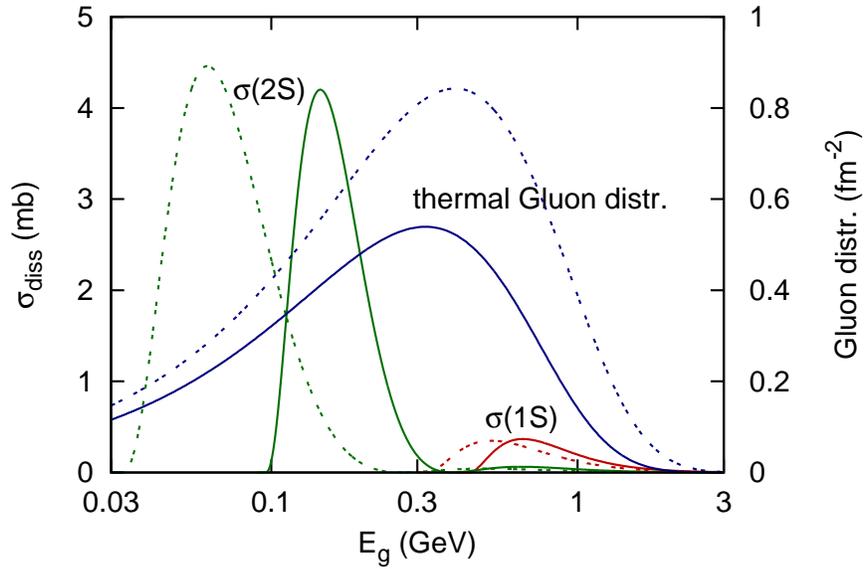}
\caption{\label{fig2} (color online) Gluodissociation cross sections $\sigma_{diss}(nS)$ in mb (lhs scale) of the  $\Upsilon (1S)$ and $\Upsilon(2S)$ states calculated using the screened Cornell potential for temperatures $T=$200 (solid curves) and 250 MeV (dotted curves) as functions of the gluon energy $E_g$. 
The thermal gluon distribution (rhs scale, with $m_g=0$; solid for $T=200$ MeV, dotted for 250 MeV)
is used to obtain the thermally averaged cross sections through integrations over the gluon momenta.}
\end{center}
\end{figure}

We obtain new results for the $2S$ and $3S$ states from eqs.~(2),(3). We also calculate the cross sections for the $\chi_b(1P)$ and $\chi_b(2P)$ states \cite{br11}. The  gluodissociation cross sections resulting from eqs.~(\ref{improvedsigma2}),(3) including the effect of screening for finite string tension are shown in Fig.~\ref{fig2} for the $1S$ and $2S$ states.

\begin{table}
\begin{center}
\caption{\label{table2}%
Calculated minimum-bias suppression factors $R_{AA}(1S)$ of the $\Upsilon (1S)$ state
for initial central temperatures $T_0$ = 500--800 MeV in 2.76 TeV PbPb,  corresponding results for an effective gluon mass of 1 GeV, and ratio of the yields $\Upsilon(2S+3S)/\Upsilon(1S)$ for effective gluon masses $m_g=0$ (fourth column), and $m_g=1$ GeV (last column). Suppression factor values for finite $m_g$ are lower bounds, see text.}
\vspace{.4cm}
\begin{tabular}{llllcr}
\hline
$T_0$(GeV)&$R_{AA} (1S)$&$R_{AA}^{m_g}(1S)$&&$\Upsilon(2S+3S)/\Upsilon(1S)$\\

\hline
     
0.8&0.50&0.61&&0.45&0.38 \\
0.7&0.53&0.66&&0.49&0.40\\
0.6&0.56&0.69&&0.53&0.44 \\
0.5&0.62&0.74&&0.58&0.50\\

\hline

\end{tabular}
\end{center}
\end{table}
One should be prepared to expect modifications in the cross section values of the five states from next-to-leading order (NLO) contributions \cite{sol05}, where a gluon appears in the final state in addition to the $b$ and $\bar{b}$ quarks, and hence, the phase space is larger than in leading order (LO). However, in \cite{gra01} it was shown that the quasi-free process that corresponds to NLO is less important than LO for temperatures $T>$270 MeV.

Whereas the heavy quarkonium is not in thermal equilibrium with the QGP, it is reasonable to assume that the medium itself is thermalized due to the short equilibration time of about 0.6 fm/$c$ \cite{won05}, at least in the transverse direction.
Hence, we integrate the gluodissociation cross sections for the $1S, 2S$ and $3S$ states over the gluon momenta $p$, weighted with the Bose-Einstein distribution function of gluons at temperature $T$ 
to obtain the average dissociation cross sections for the $nS$ states
\begin{equation}\label{eq:avcross}
<\sigma_{{diss}}^{nS}>=\frac{g_d}{2\pi^2n_g}\int_0^\infty \sigma_{{diss}}^{nS}(E)\;\frac{p^2dp}{\exp{[E(p)/T]}-1}
\end{equation}
with $E(p)=(p^2+m_g^2)^{1/2}$, the gluon degeneracy $g_d$=16, and the gluon density as the integral over the distribution function,  $n_g=g_d T^3 \zeta(3)/\pi^2$ for $m_g=0$. Values for the thermal gluon density  at temperatures 170, 200, 300 and 400 MeV and $m_g=0$ are $n_g=$ 1.25, 2.03, 6.85 and 16.23 fm$^{-3}$, respectively.
The distribution function is shown in Fig.~\ref{fig2} (rhs scale).

The on-shell gluon energy $(p^2+m_g^2)^{1/2}$ is usually calculated assuming vanishing gluon mass $m_g=0$, but we shall also investigate the effect of a finite effective gluon mass, as has been suggested in quasi-particle models \cite{plu11} based on lattice QCD results \cite{bor10,che10}, with $m_g \simeq 0.5-1$ GeV. It is argued in \cite{blge11} that the effective mass of the gluons may initially be of the order of the gluon saturation scale, $m_g \simeq Q_s$, which is about 1 GeV at $x_{{Bjorken}}=0.01$. Here we consider an effect of a finite gluon mass only on the thermal gluon distribution, not explicitly in the cross section. The resulting average cross section values for finite $m_g$ are thus upper limits since a finite gluon mass reduces the relative velocity between the $\Upsilon$ and the gluon. 
Results for the average gluodissociation cross section in mb for $T=170-400$ MeV, $m_g=0$ are shown
in Table~\ref{table1}.


The dissociation widths $\Gamma(nS)$ of the $nS$ states are then obtained by multiplying the average cross sections with the gluon density, $\Gamma(nS)=n_g\cdot<\sigma_{diss}^{nS}>$, and similarly for the $\chi_b$ states. To compare with minimum-bias data, it is essential to consider the impact-parameter dependence. We assume a monotonic relation of the initial temperature $T_i(b)=T_0 (1-b^2/b_{cr}^2)$ on impact parameter up to a critical value $b_{cr}$ where $T_i(b_{cr})=T_{cr}=170$ MeV, with $T_0 \gtrsim 500$ MeV, and no gluodissociation beyond $b_{cr}$, to obtain the temperature-dependent suppression factor $\hat{R}(nS,b)$ (and analogously for the $\chi_b(nP)$ states) prior to feed-down at impact parameter $b$ as
\begin{equation}
\hat{R}(nS,b)=\Theta[T_i(b)-T_c]\exp{[-\Gamma(nS,b)\tau(b)] }+\Theta[T_c-T_i(b)].
\end{equation}
Here we have used an interaction time of $\tau_{max} \simeq 5-8$ fm/c in central collisions in accordance with hydrodynamic \cite{shko11} and transport \cite{zra11} approaches, with a monotonic dependence on impact parameter $\tau(b)=\tau_{max}(1-b/b_{max})$, and $b_{max}$=14.22 fm in PbPb. The numerical values shown in this Letter for minimum-bias collisions are obtained after impact-parameter averaging, for $\tau_{max}=8$ fm/c.
 
This schematic calculation accounts for the essential features of the centrality on the widths, although a fully time-dependent approach such as performed in \cite{str11,shko11,zra11,blo89,liu11}  for the destruction of the $J/\Psi$ or $\Upsilon$ meson may yield slightly modified results. It should be noted, however, that the gluodissociation widths
and in particular, $\Gamma(1S)$ have only a relatively weak temperature dependence at high temperatures where gluodissociation is relevant, such that at a given impact parameter one should not expect substantial modifications from a explicit consideration of the dynamics. 

To determine the initial bottomium population vector of all five states considered in the cascade calculation, we consider $\chi_b$ populations estimated from the CDF feed-down results 
\cite{aff00} 
0.27$\pm$0.07(stat)$\pm$0.04(sys) for $\chi_b(1P)\rightarrow \Upsilon(1S)$, 0.11$\pm$0.04(stat)$\pm$0.01(sys) for $\chi_b(2P)\rightarrow \Upsilon(1S)$;
50.9 \% of the $1S$ state is directly produced. 

With decay rates for the $nS$ states from the particle data group -- including the effect of different branching ratios into the $\mu^+\mu^-$ detection channel for $n=$1, 2, 3 --, we calculate a decay cascade that matches the final populations measured by CMS for $pp$ at 2.76 TeV \cite{cha11}, and thus provides initial populations which we use for the PbPb in-medium calculation at the same energy. Following the consideration of screening and gluodissociation of the five states, we calculate the radiative feed-down cascade in the medium for those states which have survived the strong-interaction processes at a given impact parameter $b$, to obtain the final yields in the presence of the QGP.

Our results for the suppression of the $\Upsilon(1S)$ state in PbPb relative to $pp$ at 2.76 TeV are shown in Table~\ref{table2}  for several initial QGP temperatures $T_0$,  $m_g=0$, and 1 GeV. Since the average cross sections for finite gluon mass are upper bounds, the corresponding suppression factor values represent lower bounds with respect to the influence of a finite gluon mass.

For initial central QGP temperatures 0.5 GeV $\le T_0 \le 0.6$ GeV, our results 
are consistent with the experimental value currently observed by CMS, $R_{AA}(1S)= 0.62\pm0.11$(stat)$\pm0.10$(sys) in minimum-bias PbPb collisions \cite{csi11,chat11} for both zero and finite effective gluon mass. The suppression factor of the ground state may, however, be further reduced by cold nuclear matter effects such as gluon shadowing and nuclear absorption.


For the excited states we take approximate values for the initial populations as obtained by CMS for $pp$ at $\sqrt{s}=2.76$ TeV \cite{cha11}, 
$\Upsilon(2S+3S)/\Upsilon(1S)|_{pp}\simeq0.78$. Our results for the corresponding population ratio  in PbPb are shown in the last two columns of Table~\ref{table2} with $m_g=0$ and 1 GeV, respectively.  
Even at very high initial central QGP temperatures $T_0$ they are larger than, but still consistent with the experimental value that is currently observed by CMS, $\Upsilon(2S+3S)/\Upsilon(1S)_{{PbPb}}=0.24+0.13/-0.12$(stat)$\pm$0.02(sys) \cite{cha11}. 

As an example, we obtain for
$T_0 \simeq 800$ MeV and finite effective gluon mass $m_g=1$ GeV a ratio of $0.38 +0.19/-0.12$. Here the estimated theoretical error bars account for the uncertainties in the input data that enter our calculation. With the presently available data, it seems not yet possible to further narrow down the QGP temperature due to the large error bars. 
Our result leaves room for additional suppression mechanisms of the excited  $\Upsilon$ states in PbPb collisions.
\begin{figure} 
\begin{center}
\includegraphics[width=12cm]{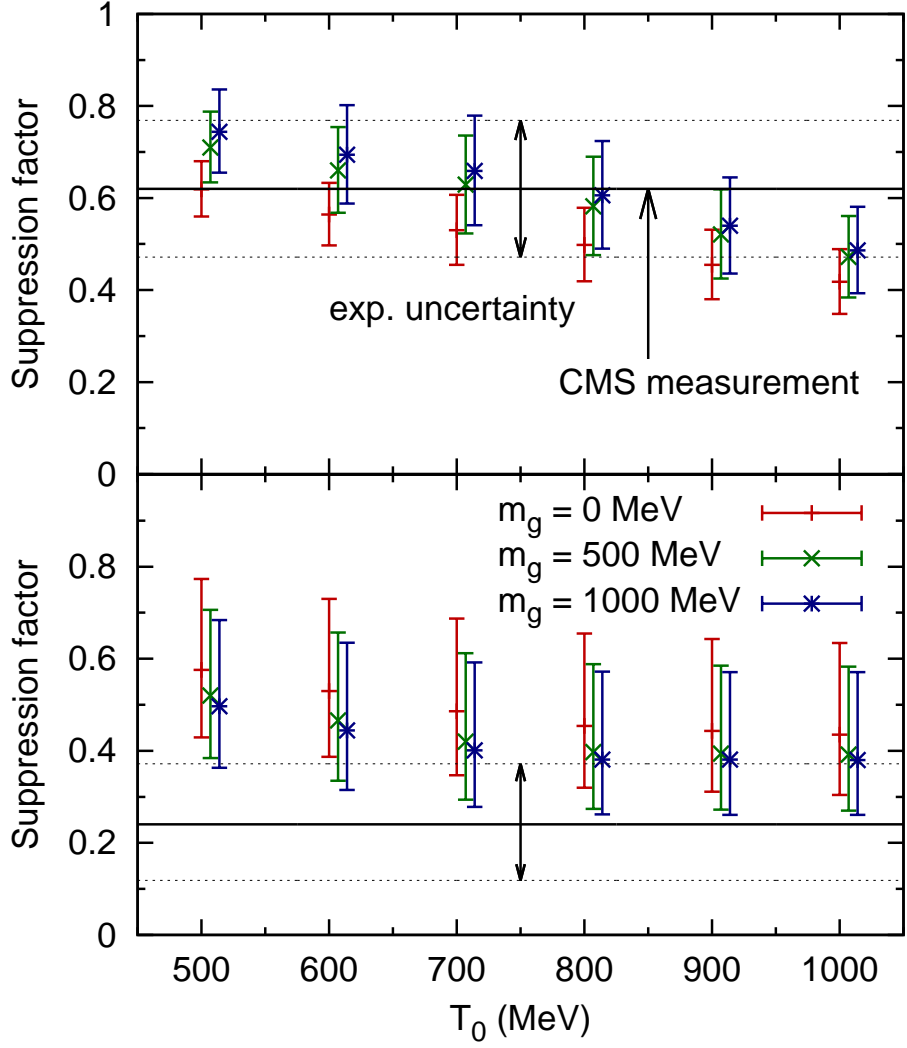}
\caption{\label{fig3} (color online) Suppression factors for the $\Upsilon(1S)$ state (top) and 
$\Upsilon(2S+3S)/\Upsilon(1S)$ (bottom) calculated in the present work for 2.76 TeV PbPb minimum-bias collisions from screening, gluodissociation and feed-down as functions of the $b=0$ temperature parameter $T_0$ for three values of the effective gluon mass $m_g$. The corresponding CMS minimum bias results (solid lines) with current statistical and systematic experimental uncertainties (dashed lines) are indicated \cite{cha11}. The estimated theoretical error bars account for the uncertainties in the input data that enter our calculation.}
\end{center}
\end{figure}

The expected physical effect, namely, rising dissociation with rising temperature, is born
out in our approach through the combination of screening, gluodissociation, and feed-down, even though the thermally averaged gluodissociation cross sections first rise and then fall with increasing temperature for the $1S$ state.
This is shown in Fig. \ref{fig3} for the $\Upsilon(1S)$ and Y(2S+3S)/1S  suppression factors in  minimum-bias collisions for three different effective gluon masses together with the CMS results \cite{cha11}, and the corresponding error bars.

To conclude, we have calculated the gluodissociation and screening of $\Upsilon (1S), (2S), (3S)$ and $\chi_b$ states at LHC energies, plus the subsequent radiative feed-down via the $\chi_b$ states. The weakly bound $3S$ state dissolves due to screening already at temperatures $T\gtrsim 200$ MeV which are close to the critical value. For $2S+3S$ relative to the $1S$ state we find a substantial suppression due to screening, gluodissociation and feed-down that is consistent with the value reported by CMS when the experimental error bars are considered, but allows for additional suppression mechanisms of the excited states.

We obtain reasonable results for the suppression of the excited $\Upsilon$ states relative to the ground state in PbPb collsions at LHC energies with an initial central QGP temperature 
of $500$ MeV$\lesssim T_0\lesssim 800$ MeV, an effective gluon mass of $m_g\simeq 0-1$ GeV, and a central-collision interaction time of $\tau_{{int}}\simeq 5-8$ fm/c. Screening and gluodissociation are relevant suppression mechanisms in particular for the higher bottomium states. The consideration of the subsequent feed-down cascade via the $\chi_b$ states turns out to be an essential ingredient in calculating the suppression of the excited states relative to the ground state. 

Although screening of the strongly bound $1S$ ground state is negligible, we find that its gluodissociation is sizeable due to the strong overlap of the $1S$ gluodissociation cross section with the thermal gluon distribution. Its observed suppression factor $R_{AA}(1S)\simeq 0.62$
in minimum-bias PbPb collisions \cite{csi11} is mainly due to both direct gluodissociation of the $1S$ state, and to the melting and gluodissociation of the $\chi_{b}(1P)$ and $\chi_{b}(2P)$ states which partially feed the $1S$ state in $pp, p\bar{p}$ and $e^+e^-$ collisions.

For a detailed comparison, one needs data with better statistics that is expected to become available from the 2011 PbPb run at the LHC. If it turned out to be possible to measure the populations of the $2S$ and $3S$ states very precisely, one could use this as a fairly accurate thermometer for the initial temperature $T_0$ of the quark-gluon plasma. On the other hand, substantial deviations from the experimental values might indicate that further mechanisms contribute to the suppression. It may, however, also turn out that the gluon distribution is not fully thermalized, in particular, in the longitudinal direction.\\\\
\bf{Acknowledgments}

\rm
This work has been supported
by the ExtreMe Matter Institute EMMI, and IMPRS-PTFS Heidelberg.
\bibliographystyle{elsarticle-num}
\bibliography{gw_up_nt}


\end{document}